\documentclass[twocolumn,10pt,cleanfoot,cleanhead]{asme2e}
\usepackage{amsmath,amsfonts,amssymb}
\usepackage{graphicx}
\usepackage[T1]{fontenc}
\usepackage[hidelinks]{hyperref}
\usepackage{mathtools}
\usepackage{xfrac}
\usepackage{multirow}
\usepackage{booktabs}
\usepackage{cite}
\graphicspath{{figures/}}

\newcommand{\bl}{\begin{lemma}}
\newcommand{\el}{\end{lemma}}
\newcommand{\be}{\begin{equation}}
\newcommand{\ee}{\end{equation}}
\newcommand{\beqn}{\begin{eqnarray}}
\newcommand{\eeqn}{\end{eqnarray}}
\newcommand{\bt}{\begin{theo}}
\newcommand{\et}{\end{theo}}
\newcommand{\bd}{\begin{df}}
\newcommand{\ed}{\end{df}}
\newcommand{\ba}{\begin{assump}}
\newcommand{\ea}{\end{assump}}
\newcommand{\bass}{\begin{assert}}
\newcommand{\eass}{\end{assert}}
\newcommand{\brem}{\begin{remark}}
\newcommand{\erem}{\end{remark}}
\newcommand{\bc}{\begin{cor}}
\newcommand{\ec}{\end{cor}}

\newcommand{\BB}{{\cal B}}

\newcommand{\pt}{\tilde{p}}

\newcommand{\St}{\tilde{S}}
\newcommand{\So}{\overline{S}}

\newcommand{\Ut}{\tilde{U}}
\newcommand{\Vt}{\tilde{V}}
\newcommand{\OBB}{\overline{\BB}}

\newcommand{\obeta}{\overline{\beta}}
\newcommand{\tbeta}{\tilde{\beta}}

\confshortname{DSCC 2018}
\conffullname{the ASME 2018 Dynamic Systems and Control Conference}
\confdate{September 30-October 3}
\confyear{2018}
\confcity{Atlanta, Georgia}
\confcountry{USA}

\papernum{DSCC2018-9125}

\title{A Dynamic-System-Based Approach to Modeling Driver Movements Across General-Purpose/Managed Lane Interfaces}

\author{Matthew A. Wright\thanks{Corresponding author.} \\
       {\tensfb Roberto Horowitz}     
    \affiliation{
	Partners for Advanced Transportation Technologies\\
	Department of Mechanical Engineering\\
	University of California\\
	Berkeley, California 94720\\
    Email: \{mwright, horowitz\}@berkeley.edu
    }	
}

\author{Alex A. Kurzhanskiy
    \affiliation{Partners for Advanced Transportation Technologies\\
	University of California\\
	Berkeley, California 94720\\
	Email: akurzhan@berkeley.edu
    }
}

\begin{document}

\maketitle

\begin{abstract}
{\it To help mitigate road congestion caused by the unrelenting growth of traffic demand, many transportation authorities have implemented managed lane policies, which restrict certain freeway lanes to certain types of vehicles. It was originally thought that managed lanes would improve the use of existing infrastructure through demand-management behaviors like carpooling, but implementations have often been characterized by unpredicted phenomena that are sometimes detrimental to system performance. The development of traffic models that can capture these sorts of behaviors is a key step for helping managed lanes deliver on their promised gains. Towards this goal, this paper presents an approach for solving for driver behavior of entering and exiting managed lanes at the macroscopic (i.e., fluid approximation of traffic) scale. Our method is inspired by recent work in extending a dynamic-system-based modeling framework from traffic behaviors on individual roads, to models at junctions, and can be considered a further extension of this dynamic-system paradigm to the route/lane choice problem. Unlike traditional route choice models that are often based on discrete-choice methods and often rely on computing and comparing drivers' estimated travel times from taking different routes, our method is agnostic to the particular choice of physical traffic model and is suited specifically towards making decisions at these interfaces using only local information. These features make it a natural drop-in component to extend existing dynamic traffic modeling methods.}
\end{abstract}



\section{INTRODUCTION}

Traffic demand in the developed and developing worlds shows no sign of decreasing, and the resulting congestion remains a costly source of inefficiency in the built environment.
One study \cite{urban_mobility_2015} estimated that, in 2014, delays due to congestion cost drivers 7 billion hours and \$160 billion in the United States alone, leading to the burning of 3 billion extra gallons of fuel.
The historical strategy for accommodating more demand has been construction of additional infrastructure, but in recent years planners have also developed strategies to improve the performance of \emph{existing} infrastructure, both through improved road operations (i.e., control strategies) and \emph{demand management}, which seeks to lower the number of vehicles on the road \cite{kurvar15}.
One such strategy that has been widely adopted in the United States and other developed countries is the creation of so-called \emph{managed lanes} \cite{obenberger_managed_lanes}.
Managed lanes are implemented on freeways by restricting the use of one or more lanes to certain vehicles.
As an example, high-occupancy-vehicle (HOV) lanes permit only vehicles with at least two (or more) occupants.
HOV lanes are intended to incentivize carpooling, which reduces the total number of cars on the road as a demand management outcome \cite{chang2008hovpolicy}.

However, the traffic-operational effects of managed lanes are not always straightforward or as rehabilitative as might be expected, as their presence can create complex traffic dynamics \cite{jangcassidy12}.
This is perhaps not too surprising: even in a freeway with simple geometry, the dynamics of traffic flow are complex, nonlinear, and not fully understood, and adding managed lanes alongside the non-managed, general-purpose (GP) lanes only exacerbates this.
Qualitatively speaking, adding a managed lane to a freeway creates two parallel and distinct, but coupled, traffic flows on the same physical structure.
When used as intended, managed lanes carry flows with different dynamics than the freeway (that is, the spatiotemporal evolution of road traffic state variables like vehicle densities and flow speeds are both qualitatively and quantitatively different between the two lane groups).
When vehicles move between the two lane flows, these two heterogeneous flows mix, and complex phenomena that are unobserved in GP-only freeways can emerge (see, e.g., \cite{jangcassidy12, daganzo_effects_2008, cassidy_smoothing_2010, liu2011friction, cassidy2015managedlanes}, and others).

Making better use of managed lanes and developing suitable traffic control strategies require an understanding of the macro-scale (i.e., network-level) behavior they induce.
One widely-used tool for understanding traffic flow behavior (and a base of many widely-used traffic control methods) is the macroscopic traffic flow model.
Macroscopic models describe traffic dynamics by approximating them as a compressible fluid.
A rich literature exists on macroscopic models for flows on long roads, and at junctions where those roads meet, but an extension to the parallel-flows situation, created by placing a managed lane in parallel and in continuous interface with a freeway, is not straightforward.

The paper addresses one such modeling concern: defining the portion of managed-lane-eligible vehicles that choose to enter or exit the managed lane, in terms of a dynamic response to current local traffic conditions.
As we will discuss below, our approach differs from traditional route- and lane-choice models in the transportation modeling literature both in assumptions and execution.
We assume no knowledge of outside traffic conditions (that is, away from the GP/managed lane interface) except the number of vehicles immediately upstream and downstream of the interface in question.
This assumption is in contrast to other lane assignment models (e.g., \cite{shiomi2015}), which model vehicles as choosing their lanes inversely proportional to, e.g., the downstream lanes' distribution of traffic speed.
As a consequence, this makes our approach agnostic to the particular form of traffic physics model used for straight roads (this point will be discussed in more detail below).
We also design our driver intentions solver as an iterative algorithm, that can be understood as a dynamic system that converges to a solution that attempts to balance the available space downstream of the interface in both lane groups.
This design choice is inspired by our recent work in modeling general traffic junctions as dynamic systems \cite{wright_node_dynamic_2016,wright2017generic}\footnote{
These dynamic system models for junctions may in turn be considered as extensions of older dynamic system characterizations of models for flows on single roads, which have been used to derive many control and optimization methods for vehicle traffic. See \cite{alinea91}, for example, for a classic paper that uses a dynamic system interpretation of a simple macroscopic traffic ``link model'' (that we will review in section~\ref{sec:background}) to design a PID-type controller for freeway onramp metering.}, and we feel could be better suited for dynamic modeling of traffic flows than static equilibrium lane-choice models.
In addition to the particular form of our algorithm, this idea of a dynamic characterization of lane choice models, as opposed to traditional statistical modeling of equilibrium conditions, is (to the best of our knowledge) novel and a contribution of this work.

The remainder of this paper is organized as follows.
Section \ref{sec:background} reviews the relevant background of traffic dynamics modeling.
Section \ref{sec:setup} presents our considered problem of driver movement intentions at GP/managed lane interfaces and gives mathematical precision to the concepts introduced in section \ref{sec:background}.
Section \ref{sec:routechoice} goes into more detail of the route or lane intentions modeling problem, discusses existing approaches, and explains our reasoning towards proposing a new approach to the specific problem of lane choice in GP/managed lane interfaces.
Section \ref{sec:solver} presents our proposed method for solving the driver movement intentions problem as well as an illustrative example.
Finally, section \ref{sec:conclusion} concludes and gives some closing remarks on how this paper's proposed idea of a dynamic-system-based lane choice model fits into the larger field of traffic dynamics modeling.

\section{BACKGROUND}
\label{sec:background}
Dynamic models for vehicular traffic flow are often differentiated by their level of abstraction.
At one end, ``microscopic'' traffic flow models consider each vehicle in isolation, and model their acceleration and turning behavior in response to the geometry of the road and other individual vehicles on it.
On the other end, ``macroscopic'' traffic flow models aggregate vehicles into a spatial and temporal continuum, and, following compressible fluid models, model the traffic flow's evolution over time as following some partial differential equation (PDE) model.
In this paper, we are concerned with traffic flow models of this macroscopic, fluid-like type.

The simplest and most-well-known PDE model for traffic is known as the ``kinematic wave'' or ``Lighthill-Whitham-Richards'' (LWR) (after \cite{Lighthill55,Richards56}) model of traffic, which states that the movement of traffic density $\rho$ through time $t$ and lineal road direction $x$ follows a simple conservation law,
\begin{equation}
	\frac{\partial \rho(x,t)}{\partial t} + \frac{\partial f(\rho, x, t)}{\partial x} = 0 \label{eq:lwr}
\end{equation}
where $f(\rho, x, t)$ is some nonlinear flux function.

In practice, macroscopic models are simulated via finite-volume approximation, similar to traditional computational fluid dynamics.
Roads are segmented into small cells of spatially-constant density, and Riemann problems are solved according to $f(\cdot)$ at each cell-cell interface, at each timestep.
A road network is modeled as a directed graph.
Each road is modeled as an edge (in the literature, typically called a link), made up of one or more of these finite-volume cells; and the junctions and interfaces where the roads (i.e., links) meet are called nodes.

Usually the flow model $f(\cdot)$ for long, straight roads is called the ``link model,'' while the $f(\cdot)$ used at nodes is called the ``node model.''
There exist many proposed link and node models in the literature that attempt to capture behaviors observed in real-world traffic.
See \cite{nie_linkmodels_2005} for an overview and comparison of some simple link models, and \cite{tampere11,wright_node_2017} for recent discussions of node models.

The finite-volume discretized form of the LWR model (Eqn. \eqref{eq:lwr}) is often called the ``cell transmission model'' (CTM) due to \cite{daganzo94}.
In the CTM, the link model $f(\cdot)$ for cells is broken into two components, one each applied at the upstream and downstream boundaries of the cell.
The function of interest at the downstream boundary of the cell is the sending function $S(\rho, x, t)$, which describes the amount of vehicles that wish to exit the cell during the current simulation timestep.
The function of interest at the upstream boundary of the cell is the receiving function $R(\rho, x, t)$, which describes the maximum amount of vehicles that the cell can accept during the current simulation timestep.
For a particular cell $\ell$, the value of the receiving function $R_\ell$ is often called the cell's supply, and the value of the sending function $S_\ell$ is often called the demand.
The node model in the CTM takes the upstream demands and downstream supplies, and returns each upstream cell's output flows and downstream cell's input flows.

In the simplest possible node where there is only one input cell and one output cell, the total flow between the two adjacent cells will be the minimum of the upstream cell's sending function and the downstream cell's receiving function.
Of course, when the node is more complex and models a merge (more than one cell upstream and a single cell downstream), a diverge (a single cell upstream and more than one cell downstream), or a merge-diverge (more than one cell both upstream and downstream), finding the flows across the node becomes more complex, and involves new modeling choices (i.e., choices as to which vehicles get to claim available space when multiple inflows are competing to claim that space).
See \cite{tampere11,wright_node_2017} for some recent overviews and discussions of the technicalities of node models.
In particular, the GP/managed lane interfaces considered in this paper take the form of a merge-diverge junction, with two cells (the GP lane and the adjacent managed lane) both entering the node and exiting the node.

In a node where there is more than one output, we model the portioning of driver intentions with \emph{split ratios} $\beta_{i,j}$, where $\beta_{i,j}$ is the portion of input link $i$'s demand that wants to go to output link $j$.
By definition, $\sum_j \beta_{i,j} = 1$.

In addition, many node models make use of some ``priority'' value to internally assign supply among the competing input links, following \cite{daganzo95a}.
We denote the priority value for link $i$ as $p_i$.
The exact form of the priority (and, if applicable, what physical quantity it is meant to describe) is a modeling choice in the node model \cite{tampere11,smits_family_2015,wright_node_2017}.
In the remainder of this paper, we assume that the node model used has some sort of priority $p_i$ to apportion downstream supply (with the supply available to link $i$ being proportional to its priority $p_i$), but do not specify its form further.

We conclude this section with one final introductory point.
In our situation where we have managed lanes that are only available to some sub-population of the traffic flow, we need to distinguish between permitted vehicles and non-permitted vehicles.
We do this by adopting a notation of different vehicle classes $c$, $c \in \{1, \dots, C\}$.\footnote{In the transportation literature, vehicle classes are sometimes called vehicle commodities.}
We define $S_i = \sum_c S_i^c$, and give each class its own set of split ratios $\beta_{i,j}^c$, with $\sum_j \beta_{i,j}^c = 1$.
The purpose of the node model in a multi-class-flow setting is to take, for input links $i$, output links $j$, and classes $c$, the values $S_i^c, R_j,$ and $\beta_{i,j}^c$ and compute the node throughflows $f_{i,j}^c$, where $f_{i,j}^c$ denotes the flow of class $c$ vehicles from link $i$ to link $j$.
In the remainder of this paper, we omit the class index $c$ when it is not relevant.

\section{PROBLEM SETUP}
\label{sec:setup}
In this section, we make precise our mathematical model for the macroscopic GP/managed lane interface.

\begin{figure}
\centering
\includegraphics{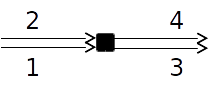}
\caption{MERGE-DIVERGE INTERSECTION EXAMPLE.}
\label{fig:node} 
\end{figure}

Consider Fig. \ref{fig:node}, which depicts a merge-diverge node with two incoming links (1 and 2) and two outgoing links (3 and 4).
In particular, this situation could model a GP/managed lane interface, with (e.g.) links 2 and 4 modeling a managed lane, and links 1 and 3 modeling the adjacent GP lane(s).

To be precise, since we expect the managed lane and GP lane(s), to behave differently (in terms of traffic speed, densities, etc.) we are breaking one road into two parallel links.

Suppose that we know the input link demands $S_1^c$ and $S_2^c$, and the output link supplies $R_3$ and $R_4$.
In this situation, a node flow model would take these quantities (along with the split ratios $\beta_{i,j}^c$) and compute the resulting throughflows $f_{i,j}^c$ (here, ${i \in \{1, 2\}}, {j \in \{3, 4\}}$).

Our problem concerns the situation where we do not know \emph{a priori} the split ratios $\beta_{i,j}^c$.
To give a specific example, suppose that the node depicted in Fig. \ref{fig:node} depicts an HOV lane in links 2 and 4 and the adjacent GP lane(s) in links 1 and 3, and that we have two vehicle classes, high-occupancy vehicles (HOVs) that can enter the managed lane, and low-occupancy vehicles (LOVs) which cannot.
By definition, we can say that ${\beta_{1,3}^L=1}$ and ${\beta_{1,4}^L=0}$ (where the superscript ${c=L}$ depicts the LOV class).
However, it is difficult to say what the split ratios for the HOVs should be.
It would make sense for the HOVs to be able to choose their lane in response to local traffic conditions.
If the portion of HOVs is low relative to the total amount of traffic, we should expect them to favor the HOV lane.
On the other hand, if the traffic population is entirely HOVs, then the HOV lane should behave in a qualitatively similar fashion to the GP lane(s) in terms of flow dynamics.

The example discussed so far in this section is the simplest possible example, with two input links, two output links, and only one class whose split ratios are undefined (we will revisit this example in section \ref{sec:example} and use the method to be discussed in section \ref{sec:solver} to solve it).
However, it is generally required that node models be applicable to arbitrary amounts of input links, output links, and vehicle classes \cite{tampere11,wright_node_2017}.
In this paper we ensure that our method for solving for driver split ratios is also generally-applicable.

\section{DRIVER ROUTE AND LANE CHOICE MODELING}
\label{sec:routechoice}
In transportation networks such as the road network, travelers usually have several routes they may choose from to reach their destination.
In the transportation literature, a very common class of route choice model is based on logit models (also known as a logit regression or logistic regression model).
Logit models for discrete choices have a long history in econometrics, where they have been used to model the probability of an individual taking one of several discrete choices \cite{mcfadden_conditional_1973}.
As applied to transportation and route choice, a logit model gives each route some relative utility as a function of relevant characteristics (travel time, monetary costs for tolls, etc.).
Then, the utilities of each choice are used to compute probabilities that an individual will take that choice, with more probability given to higher-utility choices (to be precise, this is done by using the choice utilities as inputs to a logistic or softmax function).

In transportation modeling, logit models have long been used for modeling choices among independent routes (see \cite{bliemer_impact_2008} for an overview of some route choice models used in the literature).
More recently, some authors have proposed using logit models to describe the portioning of vehicles across lanes \cite{farhi2013logit,shiomi2015}.
In logit models for lane portioning, the utility functions for lanes are often chosen to be nondecreasing functions of the speed of traffic in each lane, and sometimes with additional utility for following rules of the road such as remaining in the non-passing lanes (e.g., the right lane in the U.S.) \cite{shiomi2015}.
However, the speed of traffic in a macroscopic traffic model, which by definition from Eqn. \ref{eq:lwr} is equal to $f(\rho)/\rho$, is directly a function of the particular link and node models chosen.
As we will discuss in the following section, we make an explicit design choice to not rely on estimates of lane speed.

\subsection{Goals of our proposed method}
In this paper, we remain agnostic as to the particular link model chosen, its sending and receiving functions $S(\cdot)$ and $R(\cdot)$, as well as the particular node model.
Therefore, our method for solving for driver intentions that will be presented in section \ref{sec:solver} uses only the supplies of the downstream cells and the demands of the upstream cells, without attempting to ``peek into'' the supply and demand functions and measure traffic speeds as might be done in a utility-function-based logit lane distribution model.
In addition, this choice makes our method explicitly node-local: like node models (that is, models for computing node flows) themselves, our proposed method is restricted to these outputs.
This choice means that our models for GP/managed lane interface split ratios does not include any extra information of the neighboring links' states than the existing node models, and can be considered generally applicable and supplemental to existing node models.

In addition, the setting of GP/managed lane interfaces is qualitatively different from traffic distributions across lanes of the same type.
For example, the GP/managed lane interface has been observed to exhibit unique phenomena, such as e.g., 1) a ``friction effect'' \cite{liu2011friction,jangcassidy12}, where vehicles in a GP lane adjacent to the managed lane(s) move slower than what would be expected based solely on their density of cars (the most common theory is that these GP lane vehicles move slowly out of fear that vehicles may suddenly move out of the managed lane in front of them), or 2) a ``smoothing effect'' at bottlenecks \cite{cassidy_smoothing_2010} that leads to an \emph{increased} flow in the GP lanes closest to the managed lane(s) by reducing lane changes (briefly, fewer vehicles being eligible for the fastest lane means that fewer vehicles will change lanes - and in the process of changing lanes, slow down surrounding traffic - to enter it).
The presence of these phenomena suggests that GP lanes adjacent to managed lanes may need more sophisticated flow models than non-adjacent GP lanes, and that using a typical lane-assignment model that relies on traffic speed may have unintended second-order modeling effects due to the divergence between the observed and the predicted GP lane speed.

The above points further motivate our choice to remain agnostic to a particular flow model.

\section{PROPOSED ``BALANCING'' SPLIT RATIO SOLVER}
\label{sec:solver}
This section fully describes our method for solving for fully- or partially-undefined split ratios.
In particular, we consider a node with $M$ input links, $N$ output links,
and $C$ vehicle classes, where some of the split ratios $\beta_{ij}^c$
are not defined \emph{a priori} and must be computed as functions
of the input demand $S_i^c$, priorities $p_i$ and the output supply $R_j$,
${i=1,\dots,M}$, ${j=1,\dots,N}$ and ${c=1,\dots,C}$.

As discussed above, it is desired that the split ratios are allowed to vary in response to local traffic conditions without using more than these values, which are traditionally the values known to a node flow model.
To do this, we suggest attempting to balance the ratios of demand to supply for link $j$,
$(\sum_i \sum_c \beta_{i,j}^c S_{i}^c) / R_j$.
This is a useful proxy for link speed because the two quantities are roughly
inversely proportional --- a link $j$, 
whose supply $R_j$ is more demanded (i.e., its ratio is large),
will generally become more congested, and thus tend to have a lower speed.

As mentioned above, our proposed algorithm is inspired by our recent efforts in recharacterizing traditional node flow models as dynamic systems \cite{wright_node_dynamic_2016,wright2017generic}, which as mentioned above were in turn were inspired by classical results based on characterizing link models as dynamic systems (e.g., \cite{alinea91}).
In the present case, we propose a controller that attempts to balance these supply-to-demand ratios by assigning the unknown split ratios $\beta_{i,j}^c$.
Its exact form is described next.

\subsection{Definitions and Assumptions}
\begin{enumerate}
\item Define the set of class movements for which split ratios
are known as $\BB=\left\{\left\{i,j,c\right\}: \; \beta_{ij}^c \in [0,1]\right\}$,
and the set of class movements for which split ratios are to be
computed as $\OBB=\left\{\left\{i,j,c\right\}: \; \beta_{ij}^c \mbox{ are unknown}\right\}$.

\item For a given input link $i$ and class $c$ such that $S_i^c=0$,
assume that all split ratios are known: $\{i,j,c\}\in\BB$.\footnote{If split
ratios were undefined in this case, they could be assigned arbitrarily.}

\item Define the set of output links for which there exist unknown
split ratios as $V=\left\{j: \; \exists \left\{i,j,c\right\}\in\OBB\right\}$.

\item Assuming that for a given input link $i$ and class $c$,
the split ratios must sum up to 1, define the unassigned portion
of flow by $\obeta_i^c=1-\sum_{j:\{i,j,c\}\in\BB}\beta_{ij}^c$.

\item For a given input link $i$ and class $c$ such that there exists
at least one class movement $\{i,j,c\}\in\OBB$, assume $\obeta_i^c>0$, otherwise the
undefined split ratios can be trivially set to 0.

\item For every output link $j\in V$, define the set of input links
that have an unassigned demand portion directed toward this output link
by $U_j=\left\{i: \; \exists\left\{i,j,c\right\}\in\OBB\right\}$.

\item For a given input link $i$ and class $c$, define the set
of output links for which split ratios for which are to be computed as
$V_i^c = \left\{j: \; \exists i\in U_j\right\}$,
and assume that if nonempty, this set contains at least two elements,
otherwise a single split ratio can be trivially set equal to $\obeta_i^c$.

\item Assume that input link priorities are nonnegative, $p_i\geq 0$,
	${i=1,\dots,M}$, and ${\sum_{i=1}^M p_i = 1}$. Note that, although in
	section~\ref{sec:background} we did not require the input priorities
	to sum to one, in this section we assume this normalization is done
	to simplify the notation.

\item Define the set of input links with zero priority:
${U_{zp} = \left\{i:\;p_i=0\right\}}$.
To enable split ratio assignment for inputs with zero priorities,
perform regularization:
\be
\pt_i = p_i\left(1-\frac{|U_{zp}|}{M}\right)+ \frac{1}{M}\frac{|U_{zp}|}{M}=
p_i\frac{M-|U_{zp}|}{M} + \frac{|U_{zp}|}{M^2}
\label{priority_regularization}
\ee
where $|U_{zp}|$ denotes the number of elements in set $U_{zp}$.
Equation~\eqref{priority_regularization} implies that the regularized
input priority $\pt_i$ consists of two parts:
(1) the original input priority $p_i$ normalized to the portion of 
input links with positive priorities; and
(2) uniform distribution among $M$ input links, $\frac{1}{M}$, 
normalized to the portion of input links with zero priorities.

Note that the regularized priorities $\pt_i> 0$, $i=1,\dots,M$, and $\sum_{i=1}^M \pt_i = 1$.
\end{enumerate}

\subsection{Algorithm}
The algorithm for distributing $\obeta_i^c$ among the class movements
in $\OBB$ (that is, assigning values to the \emph{a priori} unknown split ratios)
aims at maintaining output links as uniform in their demand-supply ratios as possible.
At each iteration $k$, two quantities are identified: $\mu^+(k)$, which is the largest \emph{oriented} demand-supply ratio produced by the split ratios that have been assigned so far, and $\mu^-(k)$, which is the smallest oriented demand-supply ratio whose input link, denoted $i^-$, still has some unclaimed split ratio.
Once these two quantities are found, the class $c^-$ in $i^-$ with the smallest unallocated demand has some of its demand directed to the $j$ corresponding to $\mu^-(k)$ to bring $\mu^-(k)$ up to $\mu^+(k)$ (or, if this is not possible due to insufficient demand, all such demand is directed).

To summarize, in each iteration $k$, the algorithm attempts to bring the smallest oriented demand-supply ratio $\mu^-(k)$ up to the largest oriented demand-supply ratio $\mu^+(k)$.
If it turns out that all such oriented demand-supply ratios become perfectly balanced, then the demand-supply ratios $(\sum_i \sum_c S_{ij}^c) / R_j$ are as well.

The algorithm is:

\begin{enumerate}
\item Initialize:
\begin{eqnarray*}
\tbeta_{ij}^c(0) & := & \left\{\begin{array}{ll}
\beta_{ij}^c, & \mbox{ if } \{i,j,c\}\in\BB,\\
0, & \mbox{ otherwise};\end{array}\right. \\
\obeta_i^c(0) & := & \obeta_i^c; \\
\Ut_j(0) & = &  U_j; \\
\Vt(0) & = & V; \\
k & := & 0,
\end{eqnarray*}
Here $\Ut_j(k)$ is the remaining set of input links with some unassigned demand,
which may be directed to output link $j$; and
$\Vt(k)$ is the remaining set of output links, to which the still-unassigned
demand may be directed.

\item If $\Vt(k)=\emptyset$, stop.
The sought-for split ratios are $\left\{\tbeta_{ij}^c(k)\right\}$,
$i=1,\dots,M$, $j=1,\dots,N$, $c=1,\dots,C$.

\item Calculate the remaining unallocated demand:
\[
\So_i^c(k) = \obeta_i^c(k) S_i^c, \;\;\; i=1,\dots,M, \;\; c=1,\dots,C.
\]

\item For all input-output link pairs,
calculate as a shorthand, the ``oriented demand'':
\[ \St_{ij}^c(k) = \tbeta_{ij}^c(k) S_i^c. \]

\item For all input-output link pairs, calculate as a shorthand the ``oriented priorities'':
\begin{equation}
\pt_{ij}(k) = \pt_i\frac{\sum_{c=1}^C \gamma_{ij}^c S_i^c}{
\sum_{c=1}^C S_i^c}
\label{eq_oriented_priorities_undefined_sr1}
\end{equation}
where
\begin{align}
\gamma_{ij}^c(k) = \begin{cases}
\beta_{ij}^c & \begin{aligned}
	& \text{if split ratio is defined} \\
	& \text{\emph{a priori}: } \{i,j,c\} \in \BB 
	\end{aligned} \\	
\tbeta_{ij}^c(k) + \frac{\obeta_i^c(k)}{|V_i^c|} & \text{otherwise}
\end{cases}
\label{eq_oriented_priorities_undefined_sr2}
\end{align}
where $|V_i^c|$ denotes the number of elements in the set $V_i^c$.
Examining Eqns.~\eqref{eq_oriented_priorities_undefined_sr1}-\eqref{eq_oriented_priorities_undefined_sr2}, one can see that the
split ratios $\tbeta_{ij}^c(k)$, which are not fully defined yet,
are complemented with a fraction of $\obeta_i^c(k)$ inversely proportional
to the number of output links among which the flow of class $c$
from input link $i$ can be distributed.

Note that in this step we are using \emph{regularized} priorities $\pt_i$
as opposed to the original $p_i$, $i=1,\dots,M$.
This is done to ensure that inputs with $p_i=0$ are not ignored
in the split ratio assignment.

\item Find the largest oriented demand-supply ratio:
\[
\mu^+(k) = \max_{j} \max_{i}
\frac{\sum_{c=1}^C \St_{ij}^c(k)}{\pt_{ij}(k)R_j}\sum_{i\in U_j}\pt_{ij}(k).
\]

\item Define the set of all output links in $\Vt(k)$, where the minimum of
the oriented demand-supply ratio is achieved:
\[
Y(k) = \arg\min_{j\in\Vt(k)}\min_{i\in\Ut_j(k)}
\frac{\sum_{c=1}^C \St_{ij}^c(k)}{\pt_{ij}(k)R_j}
\sum_{i\in U_j}\pt_{ij}(k),
\]
and from this set pick the output link $j^-$ with the smallest
output demand-supply ratio (when there are multiple
minimizing output links, any of the minimizing output links
may be chosen as $j^-$):
\[
j^- = \arg\min_{j\in Y(k)}
\frac{\sum_{i=1}^M\sum_{c=1}^C\St_{ij}^c(k)}{R_j}.
\]

\item Define the set of all input links, where the minimum of
the oriented demand-supply ratio for the output link $j^-$ is achieved:
\[
W_{j^-}(k) = \arg\min_{i\in\Ut_{j^-}(k)}
\frac{\sum_{c=1}^C \St_{ij^-}^c(k)}{\pt_{ij^-}(k)R_{j^-}}
\sum_{i\in U_{j^-}}\pt_{ij^-}(k),
\]
and from this set pick the input link $i^-$ and class $c^-$
with the smallest remaining unallocated demand:
\[
\{i^-, c^-\} = \arg\min_{\begin{array}{c}
i\in W_{j^-}(k),\\
c:\obeta_{i^-}^c(k)>0\end{array}} \So_i^c(k).
\]

\item Define the smallest oriented demand-supply ratio:
\begin{equation*}
\mu^-(k) = 
\frac{\sum_{c=1}^C \St_{i^-j^-}^c(k)}{\pt_{i^-j^-}(k)R_{j^-}}
\sum_{i\in U_{j^-}}\pt_{ij-}(k).
\end{equation*}
\begin{enumerate}
\item If $\mu^-(k) = \mu^+(k)$, the oriented demands created by
the split ratios that have been assigned as of iteration $k$,
$\tbeta_{ij}^c(k)$, are perfectly balanced among the output links,
and to maintain this, all remaining unassigned split ratios should
be distributed proportionally to the allocated supply:
\begin{align}
\begin{split}
\tbeta_{ij}^{c}(k+1) &= 
\begin{aligned}[t]
&\tbeta_{ij}^{c}(k) + 
\frac{R_j}{\sum_{j'\in V_{i}^{c}(k)}R_{j'}}
\obeta_{i}^{c}(k), \\ 
&c: \obeta_i^c(k) > 0, \; i \in \Ut_j(k), \; j\in \Vt(k); \label{eq_sr_assign_1}
\end{aligned}
\end{split} \\
\obeta_{i}^{c}(k+1) &= 0, \;\;\; c:~\obeta_i^c(k) > 0,
\;\; i \in \Ut_j(k), \;\; j\in \Vt(k); \nonumber\\
\Ut_j(k+1) &= \emptyset, \;\;\; j\in\Vt(k); \nonumber \\
\Vt(k+1) &= \emptyset. \nonumber
\end{align}
If the algorithm ends up at this point, we have emptied $\Vt(k+1)$ and are done.
\item Else, assign:
\begin{align}
\begin{split}
\Delta\tbeta_{i^-j^-}^{c^-}(k) &= \min \Bigg\{ \obeta_{i^-}^{c^-}(k), \\
	& \hspace{-40pt}\Bigg( \frac{\mu^+(k)\pt_{i^-j^-}(k)R_{j^-}}{
	\So_{i^-}^{c^-}(k) \sum_{i\in U_{j^-}}\pt_{ij^-}(k)} -
	\frac{\sum_{c=1}^C\St_{i^-j^-}^c(k)}{\So_{i^-}^{c^-}(k)}\Bigg)\Bigg\}
	\label{eq_sr_assign_3}
	\end{split}\\
\tbeta_{i^-j^-}^{c^-}(k+1) & = \tbeta_{i^-j^-}^{c^-}(k) + 
\Delta\tbeta_{i^-j^-}^{c^-}(k);
\label{eq_sr_assign_4}\\
\obeta_{i^-}^{c^-}(k+1) & = \obeta_{i^-}^{c^-}(k) -
\Delta\tbeta_{i^-j^-}^{c^-}(k); \label{eq_sr_assign_5} \\
\tbeta_{ij}^{c}(k+1) & = \tbeta_{ij}^{c}(k) \mbox{ for }
\{i,j,c\}\neq\{i^-,j^-,c^-\}; \nonumber\\
\obeta_i^c(k+1) & = \obeta_i^c(k) \mbox{ for } \{i,c\}\neq\{i^-,c^-\};
\nonumber \\
\Ut_j(k+1) & = \Ut_j(k) \setminus 
	\begin{aligned}[t]
		\Big\{&i: \obeta_{i}^c(k+1) = 0, \\
		&c=1,\dots,C \Big\}, \;\;\; j\in \Vt(k); 
	\end{aligned} \nonumber\\
\Vt(k+1) & = \Vt(k) \setminus \left\{j: \; \Ut_j(k+1)=\emptyset\right\}.
\nonumber
\end{align}
In Eqn.~\eqref{eq_sr_assign_3}, we take the minimum of the remaining unassigned
split ratio portion $\obeta_{i^-}^{c^-}(k)$ and the split ratio portion
needed to equalize $\mu^-(k)$ and $\mu^+(k)$.
To better understand the latter, the second term in the $\min\{\cdot,\cdot\}$ can
be rewritten as
\begin{multline*}
\frac{\mu^+(k)\pt_{i^-j^-}(k)R_{j^-}}{
\So_{i^-}^{c^-}(k) \sum_{i\in U_{j^-}}\pt_{ij^-}(k)} -
\frac{\sum_{c=1}^C\St_{i^-j^-}^c(k)}{\So_{i^-}^{c^-}(k)} \\
 =
\left(\frac{\mu^+(k)}{\mu^-(k)}-1\right)
\left(\sum_{c=1}^C\St_{i^-j^-}^c(k)\right)
\frac{1}{\So_{i^-}^{c^-}(k)}.
\end{multline*}
The right hand side of the last equality can be interpreted as:
flow that must be assigned for input $i^-$, output $j^-$ and class $c^-$
to equalize $\mu^-(k)$ and $\mu^+(k)$ minus flow that is already assigned
for $\{i^-,j^-,c^-\}$, divided by the remaining unassigned portion
of demand of class $c^-$ coming from input link $i^-$.

In~Eqn. \eqref{eq_sr_assign_4} and~Eqn. \eqref{eq_sr_assign_5}, the assigned
split ratio portion is incremented and the unassigned split ratio portion is
decremented by the computed $\Delta\tbeta_{i^-j^-}^{c^-}(k)$.
\end{enumerate}

\item Set $k := k+1$ and return to step 2.
\end{enumerate}

\subsection{Example}
\label{sec:example}
Recall again the example node depicted in Fig. \ref{fig:node}.
As a brief recap, we will say that links 1 and 3 represent a freeway's GP lanes, and links 2 and 4 represent an HOV
lane, open only to a select group of vehicles. The Low-Occupancy
Vehicles (LOVs) and HOVs will be modeled as separate vehicle classes, with
notations of $L$ for LOVs and $H$ for HOVs.

\begin{table}[t]
\caption{PARAMETERS FOR EXAMPLE IN SECTION \ref{sec:example}.}
\begin{center}
\label{tab:params}
\begin{tabular}{l l l l}
\toprule
 LOV Splits & $\beta_{14}^L = 0$ & $\beta_{13}^L = 1$ & \\
 \midrule
 \multirow{2}{*}{HOV Splits } & $\beta_{14}^H = ?$ & $\beta_{13}^H = ?$ & \\
 & $\beta_{23}^H = ?$ & $\beta_{24}^H = ?$ & \\
 \midrule
 \multirow{3}{*}{Node Model Information} & $S_1^L = 500$ & $S_1^H = 100$ & $R_3 = 600$ \\
 & $S_2^H = 50$ & & $R_4 = 200$ \\
 & $p_1=\sfrac{3}{4}$ & $p_2 = \sfrac{1}{4}$ & \\
\bottomrule
\end{tabular}
\end{center}
\end{table}
Suppose we have as input parameters the values given in Tab. \ref{tab:params}.

In words, the LOVs are not allowed to enter the HOV lane; HOVs are allowed,
but individual vehicles may or may not choose to do so: both links can
bring them to their eventual downstream destination so the choice of link
to take will be made in response to immediate local congestion conditions.

Let us outline how the algorithm would assign split ratios for the HOVs.

\underline{$k=0$}:
		$\begin{aligned}[t]
			\qquad \Delta \tbeta_{i^- j^-}^{c^-}(0) &= \Delta \tbeta_{24}^H(0)
				= 1 \\
			\tbeta_{24}^H(1) &= 0 + 1 = 1 \\
			\obeta_{2}^H(1) &= 1 - 0 = 0
		\end{aligned}$

\vspace{10pt}
\underline{$k=1$}:
$\begin{aligned}[t]
				\qquad \Delta \tbeta_{i^- j^-}^{c^-}(1) &= \Delta \tbeta_{14}^H(1)
					= \sfrac{1}{3} \\
				\tbeta_{14}^H(2) &= 0 + \sfrac{1}{3} = \sfrac{1}{3} \\
				\obeta_{2}^H(2) &= 1 - \sfrac{1}{3} = \sfrac{2}{3}
			\end{aligned}$

\vspace{10pt}
\underline{$k=2$}:
$\begin{aligned}[t]
				\qquad \tbeta_{13}^H(3) &= 0.64 \\
				\tbeta_{23}^H(3) &= 0.36 \\
				\obeta_{1}^H(3) &= 0
			\end{aligned}$

\vspace{10pt}
\underline{$k=3$}:
The algorithm terminates. The resulting split ratios are:
$\boldsymbol{ \beta_{13}^H = 0.64, \quad
	\beta_{23}^H = 0.36, \quad \beta_{24}^H = 1,
	\quad \beta_{23}^H = 0}$.

\subsubsection{Discussion.}
Note that we have \emph{not} determined how many of each vehicle class may actually be able to reach their intended destination.
We have only determined that 36\% of the HOV drivers in the GP lane will want to enter the HOV lane; after that point, the supply/demand ratios will be equalized, and there will not be a gain in relative open space for changing.
However, it is possible that \emph{not all} of these drivers who wish to change lanes will be able to do so.
The actual resulting flows will be computed by the node model, which is another modeling choice.
In fact, one reasonable modeling choice may be to state that a car in the act of changing lanes will take up space in both lanes at once for at least some amount of time.
It has indeed been argued (e.g., \cite{jin_kinematic_2010}) that areas of intense aggregate lane-changing can lead to creation of bottlenecks.
The extension of current node models to include lane-changing effects, especially in situations of potentially intense lane-changing such as GP/managed lane interfaces, is an open question.

\section{CONCLUSION}
\label{sec:conclusion}
This paper presented a new approach towards determining driver intentions in response to dynamically-changing traffic conditions.
As we have mentioned, this method is qualitatively different from existing methods in the traffic dynamics modeling literature, being based on a local view of the traffic conditions at a particular interface in a finite-volume simulation.
We feel that this particular trait makes it suited for ``smaller scale'' route choice problems, such as the motivating GP/managed lane interfaces.

It is worth taking a moment to view where this dynamic-system-based driver intention model fits in to the broader setting of macroscopic traffic dynamics modeling.
We mentioned in our background discussions that macroscopic simulations of traffic are typically thought of consisting of a ``link model'' for the dynamics of traffic on a single road, and a ``node model'' for the dynamics at junctions and intersections.
As we have also mentioned, the particular choice of link model and node model are critical modeling decisions that determine the broader network dynamics that can emerge in a road network system.
We can view the driver intentions model developed in this paper as a new vector for making modeling choices for road network dynamics.
The driver intentions (parameterized through, e.g., the split ratios described in this paper) have influence on the emergent network dynamics, just as the link and node models do.

We have suggested that our driver intentions model could be seen as an extension or component of a node model (in the traffic modeling literature, dynamics of moving between roads are captured in the node model), but analyses of node models tend to treat driver intentions (in, e.g., the form of split ratios) as exogenous (see, e.g., \cite{tampere11,wright_node_2017,jabari_node_2016}).
This assumption may be less true today than it had been in the past, as driver use of real-time traffic advisories through smartphone and in-car software have become ubiquitous.
Studying and modeling the dynamics of driver self-routing and self-re-routing in response to local conditions may be a crucial, yet unexplored, avenue to understanding modern traffic dynamics.
In turn, understanding this contributor to traffic dynamics in dynamic systems terms may enable a new vector for control of macro-scale traffic flows, as these same technologies grant controllers the ability to affect these previously-purely exogenous driver intentions.

\bibliographystyle{asmems4}

\begin{acknowledgment}
This research was supported by the California Department of Transportation.
An earlier version of this work appeared in the technical report \cite{hot_modeling_report}.
\end{acknowledgment}

%
% ...

\bibliography{traffic}

\end{document}